\documentclass [twocolumn,
amsymb,amsmath,nobibnotes,superscriptaddress,aps,prb]{revtex4}
 \usepackage {bm}
 \usepackage{graphicx}
 \usepackage{amsmath}

\begin{document}
\title{Edge superconducting state in attractive U Kane-Mele-Hubbard model}
 \author{Jie Yuan}
 \affiliation{Department of Physics, and Center of Theoretical and Computational Physics, The University of Hong Kong,
 Hong Kong, China}
\author{Jin-Hua Gao}
\email{jhgao1980@gmail.com}
 \affiliation{Department of Physics, and Center of Theoretical and Computational Physics, The University of Hong Kong,
 Hong Kong, China}
 \author{Wei-Qiang Chen}
 \email{chen.wq@sustc.edu.cn}
 \affiliation{Department of Physics, South University of Science and Technology of China, Shenzhen, Guangdong, China}
 \affiliation{Department of Physics, and Center of Theoretical and Computational Physics, The University of Hong Kong,
 Hong Kong, China}
\author{Yi Zhou}
 \affiliation{Department of Physics, Zhejiang University, Hangzhou,  China}
 \author{Fu-Chun Zhang}
 \affiliation{Department of Physics, and Center of Theoretical and Computational Physics, The University of Hong Kong,
 Hong Kong, China}
\affiliation{Department of Physics, Zhejiang University, Hangzhou,
China}
 \begin{abstract}
 We theoretically investigate the  phase transition from topological
insulator (TI) to superconductor in the attractive U Kane-Mele-Hubbard
model with self-consistent mean field method. We demonstrate the existence of edge superconducting state (ESS), in which the
bulk is still an insulator and the superconductivity only appears
near the edges. The ESS results from the special energy dispersion of TI, and is a general property of the superconductivity in TI.
The phase transition in this model essentially consists of two steps. When the attractive U becomes nonzero, ESS appears immediately. After the attractive
U exceeds a critical value $U_c$, the whole system becomes a superconductor.
The effective model of the ESS has also been discussed and we believe that the conception of ESS can be realized in atomic optical lattice system.
\end{abstract}

\maketitle

The topological insulator (TI) has drawn a great deal of attention
recently because it offers us a novel quantum state of electrons,
i.e. topological insulating state which is insulating in the bulk
but has edge states protected by time reversal symmetry \cite{kane1,moore1,sczhang1}. The topological
insulating state results from its nontrival band topology induced
by spin-orbit interaction and time reversal symmetry, and is characterized
by a $Z_{2}$ topological invariant. The experimental discovery of
two-dimensional (2D) and three-dimensional (3D) TI in a variety of
materials has greatly promoted the research interests in TI\cite{2dti,3dti1,3dti2}. Many
intriguing issues about TI has been proposed, for example, the realization
of Majarona fermion in TI\cite{kanemaja,linder}, new kinds of spintronic or magnetoelectric
device\cite{mfranz,nagaosa,shankar}, and the strong correlation effects in TI\cite{assaaad,cjwu,dhlee,fiete,imada,kyyang,lehur,lehur2,pesin,raghu,xcxie,yran}.

The TI in strong correlation system mainly concerns two kinds of
problem. One is about the electron-electron interaction induced topological
insulating state\cite{raghu,yran,kyyang}, and the other is about the novel phase transition
between the topological insulating state and strong correlation quantum
states\cite{assaaad,cjwu,dhlee,fiete,imada,lehur,lehur2,pesin,xcxie}. Kane-Mele-Hubbard model is the simplest and very important
2D strong correlation TI theoretical model\cite{assaaad,cjwu,fiete,imada,lehur,lehur2,xcxie,dhlee}. It basically is the Hubbard
model on honeycomb lattice with spin-orbit coupling. The exotic strong
correlation quantum states of the Hubbard model on honeycomb lattice,
especially the spin liquid state, has been entensively studied recently\cite{zymeng}.
Meanwhile, the Kane-Mele model\cite{kane3}, i.e. honeycomb lattice with spin-orbit
coupling, describes the 2D topological insulating state, which is also
named quantum spin hall (QSH) state  due to the analogy to the quantum
hall effect. An interesting problem is what the effect of the interplay
between the spin-orbit interaction and electron-electron interaction is.
Recently, several works have been done to discuss the phase diagram
of the Kane-Mele-Hubbard model\cite{assaaad,cjwu,fiete,imada,lehur,lehur2,xcxie,dhlee}, including Quantum Monte Carlo calculations\cite{xcxie,assaaad,cjwu,imada}.
Besides the theoretical interests, the QSH state has been proposed to be
realized in various materials\cite{ygyao,shitade,ruqianwu}.

Superconductivity in TI is also a research focus. On one hand, it
is predicted that Majarona fermion can be realized on TI surface via
inducing superconductivity by proximity effect. On the other, in experiment,
3D TI can be tuned into superconductor through doping with copper\cite{hor,wray}
or applying high pressure\cite{jlzhang,czhang}. Several theoretical works have been done
to analyze the superconductivity in these TI materials\cite{lfu,sasaki,sato,tklee}. But these investigations are
all about the 3D TI, and little attention has been paid to the superconductivity
in QSH system (2D TI). More important, all these theoretical works start
with the assumption of finite bulk superconductivity, which is suggested by experiments. But according to our knowledge, no discussion about the  quantum phase transition  between the topological insulating state and superconducting state
in TI has been presented so far,  which is surely an intriguing
theoretical issue.

In this paper, we theoretically study the influence of the attractive interaction on the topological insulating state
in the QSH
system via self-consistent mean field method. We start with the
Kane-Mele model of QSH system. The superconductivity is brought on
by the attractive Hubbard U interaction. With self-consistent mean
field method, we investigate the phase transition from the topological insulating state to the superconducting state.
Our results clearly show the existence
of the edge superconducting state (ESS), in which the bulk is still
insulating and superconductivity only appears near the edges. We
point out that the phase transition from TI into superconducting
state actually splits into two steps. Turning on the attractive U,
TI will first evolve into the ESS immediately, and after exceeding a
critical $U_c$ the whole system will become a superconductor. The
reason is straightforward. TI has a bulk gap and gapless edge
states. Electron pairs appear in the bulk only if the binding energy
is larger than the bulk gap. From this point of view, the ESS seems
to be a general property of the superconductivity in TI.
Interestingly even if the bulk gap is zero, the ESS still occurs in the half filling case
because of the special density of states of the Dirac fermion in 1D
(edge state) and 2D (bulk). We  further deduce that the ESS also
exists in 3D TI, but it has a critical $U_{edge}$ at half filling because of the
special 2D density of states. Finally, we give a short discussion
about the effective model of ESS.

 Of course, we notice that the realistic phase transition phenomena in the attractive U Kane-Mele-Hubbard model will
 be more complicated. For example, the possible CDW phase, the possible topological superconductor phase with p wave pairing, and the BCS-BEC crossover are not involved in this work.
 But in this work we  want to focus on the ESS which is a characteristic of the superconductivity of TI and has not been noticed in former studies. We believe that our mean field study offers a good and reasonable starting point for further investigation.

\begin{figure}
\centering
\includegraphics[width=8.5cm]{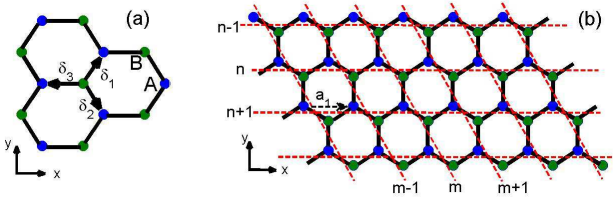}
\caption{(Color online) Schematic illustration of the honeycomb lattice (a) and Zigzag ribbon structure (b) of the QSH system.} \label{zigzagribbon}
\end  {figure}

The attractive U Kane-Mele-Hubbard model is defined as $H=H_{0}+H_{U}$,
where $H_{0}$ is just the Kane-Mele model for the QSH system and $H_{U}$
describes the attractive Hubbard U term which induces the superconductivity.
We have
\begin{equation}
\begin{split}
H_{0}=&-t\sum_{ \substack{ <ij> \\ \sigma } } c_{i\sigma}^{+}c_{j\sigma}+i\lambda_{so}\sum_{ \substack{ <<ij>>  \\
 \sigma\sigma'}}v_{ij}c_{i\sigma}^{+}\tau_{\sigma\sigma'}^{z}c_{j\sigma'} \\ &-\mu\sum_{i\sigma}c_{i\sigma}^{+}c_{i\sigma}
\end{split}
\end{equation}
where $<>$ indicates the nearest neigborhood (NN) hopping, $t$
is the hopping amplitude and $\mu$ is the chemical potential. The
second term is the spin-orbit interaction. $\lambda_{so}$ is the
spin-orbit coupling constant, and $<<>>$ denotes
the next nearest neighborhood (NNN) hopping. $\tau$ is the Pauli
matrix. $v_{ij}=\pm1$ depending on the relative orientation of the
two bonds connecting $i$ $j$ sites\cite{kane3}. Since the honeycomb lattice
is a bipartite, it is convenient to split the lattice into two sublattice
A and B. In the following, we use $a_{i\sigma}^{+}$ ($b_{i\sigma}^{+}$) denotes the creation
operator in sublattice A (B).
The Hubbard U term is $
 H_{U}=-U\sum_{i}n_{i \uparrow}n_{i \downarrow}
$
which is treated with mean field approximation ($U>0$).
In order to understand the superconducting states in QSH system well,
we first investigate the bulk properties in k space.
Then, we consider the zigzag ribbon structure which concentrates on the edge states.
 Finally, we  discuss  the effective model of the ESS.

\begin{figure}
\centering
\includegraphics[width=8.5cm]{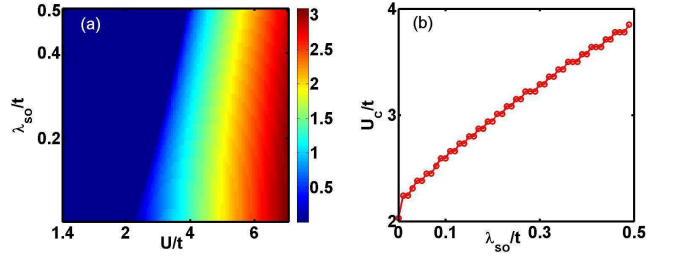}
\caption{(Color online) (a)  The zero temperature superconducting gap $\Delta(U,\lambda_{so})$ in bulk for the half filling case. (b) $U_c$ in bulk as function of $\lambda_{so}$ at half filling.} \label{zigzagribbon}
\end  {figure}

With the relation $c_{k\sigma}^{+}=\frac{1}{\sqrt{N_{A}}}\sum_{i\in A}e^{ik \cdot R_{i}}c_{i\sigma}^{+}$,
the Hamiltonian can be changed into momentum space. $N_{A}$ is the
site number of sublattice A, and sublattice A and B are equivalent.
The NN hopping is $H_{NN}=-t\sum_{k\sigma}(\gamma_{k}a_{k\sigma}^{+}b_{k\sigma}+\gamma_{k}^{*}b_{k\sigma}^{+}a_{k\sigma})$,
where $\gamma_{k}\equiv\gamma(k)=\sum_{\delta_{i}}e^{ik\delta_{i}}$.
$\delta_{i=1,2,3}$ is the nearest neighborhood vectors [See Fig. 1 (a)].
For honeycomb lattice, $\delta_{1}=a_0(\frac{1}{2},\frac{\sqrt{3}}{2})$, $\delta_{2}=a_0(\frac{1}{2},-\frac{\sqrt{3}}{2})$
and $\delta_{3}=a_0(-1,0)$ . Here $a_0$ is the lattice constant.
The spin-orbit coupling is $H_{so}=\sum_{k}\lambda_{k}[a_{k\uparrow}^{+}a_{k\uparrow}-a_{k\downarrow}^{+}a_{k\downarrow}-b_{k\uparrow}^{+}b_{k\uparrow}+b_{k\downarrow}^{+}b_{k\downarrow}]$
where $\lambda_{k}=2\lambda_{so}[-sin(\sqrt{3}a_0k_{y})+2cos(\frac{3}{2}a_0k_{x})sin(\frac{\sqrt{3}}{2}a_0k_{y})]$.
With mean field approximation, the negative Hubbard U term is $H_{U}=-\sum_{k}[\Delta_{A}a_{k\uparrow}^{+}a_{-k\downarrow}^{+}+\Delta_{A}^{*}a_{-k\downarrow}a_{k\uparrow}+\Delta_{B}b_{k\uparrow}^{+}b_{-k\downarrow}
^{+}+\Delta_{B}^{*}b_{-k\downarrow}b_{k\uparrow}]$, with $\Delta_{A}=\frac{U}{N_{A}}\sum_{k}\left\langle a_{-k\downarrow}a_{k\uparrow}\right\rangle $.
Since the sublattice A and B are equivalent, we assume $\Delta_{A}=\Delta_{B}=\Delta$. In Nambu basis $\Phi_{k}^{+}=(a_{k\uparrow}^{+},b_{k\uparrow}^{+},a_{-k\downarrow},b_{-k\downarrow})$,
we have $H=\sum_{k}\Phi_{k}^{+}H_{k}\Phi_{k}$ with \[
H_{k}=\left(\begin{array}{cccc}
\lambda_{k}-\mu & -t\gamma_{k} & -\Delta & 0\\
-t\gamma_{k}^{*} & -\lambda_{k}-\mu & 0 & -\Delta\\
-\Delta^{*} & 0 & -\lambda_{k}+\mu & t\gamma_{k}\\
0 & -\Delta^{*} & t\gamma_{k}^{*} & \lambda_{k}+\mu\end{array}\right)\]
Diagonalizing $H_{k},$ the excitation energy of the Bogoliubov quasiparticles
are $E_{\nu=\pm}(k)=\sqrt{(\epsilon_{k}\pm\mu)^{2}+\Delta^{2}}$ with
$\epsilon_{k}=\sqrt{\lambda_{k}^{2}+t^{2}|\gamma_{k}|^{2}}$ and $\nu$
is the band index. The gap equation is
\begin{equation}
\frac{1}{U}=\frac{1}{4N_{A}}\sum_{k\nu}\frac{1}{E_{\nu}}tanh(\frac{\beta E_{\nu}}{2}).
\end{equation}
The average electron density is
\begin{equation}
n-1=-\frac{1}{N_{A}}\sum_{k\nu}[\frac{\eta_\nu \epsilon_{k}-\mu}{E_{\nu}}tanh(\frac{\beta E_{\nu}}{2})]
\end{equation}
where $n=N_{e}/2N_{A}$ and $\eta_{\nu=\pm}=\pm 1$.
Thus, given $n$ and $U$, $\Delta$ and $\mu$ can be determined self-consistently with above equations\cite{cutoff}.
Here we calculate the half filling case ($n=1$) at zero temperature. The results are shown in Fig. 2.
In Fig.2 (a), we calculate the zero temperature superconducting gap $\Delta$ as a function of $U$ and $\lambda_{so}$. We see that for any $\lambda_{so}$ there exists a critical $U_c$. Only when $U>U_c$, $\Delta$ becomes nonzero. The $U_c$ here actually reflect the competition between the formation of cooper pair and the bulk gap. Since there is a bulk gap in topological insulating state, there are no free carriers to start from. The attractive U can induce superconductivity only if the binding energy is larger than the cost to produce free electrons and holes across the bulk gap. The superconductivity in such gapped system has been carefully discussed in semiconductor system\cite{nozi}. The $U_c$ is given in Fig. 2 (b) as a function of $\lambda_{so}$. We know that $\lambda_{so}$ determine the bulk gap in Kane-Mele model, where the larger $\lambda_{so}$ the larger gap is. So $U_c$ should increase with $\lambda_{so}$ since larger $U_c$ is needed to overcome the larger bulk gap. An interesting case is $U_c\approx2.1t$ when $\lambda_{so}=0$. It indicates that even without bulk gap $U_c$ still exists at half filling. This phenomenon has been noticed in former study in optical lattice system\cite{ezhao}. It results from the special density of states of the honeycomb lattice which will approach zero around the Dirac points.
\begin{figure}
\centering
\includegraphics[width=8cm]{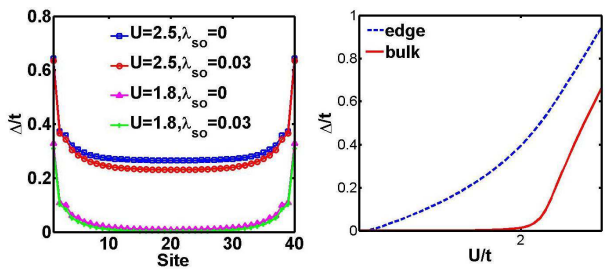}%
\caption{(Color online) (a) The zero temperature superconducting gap $\Delta$ across the zigzag ribbon at half filling ($N=20$, i.e. 40 sites across the ribbon). (b) Dashed line (blue): $\Delta (U)$ at the ribbon edge. Solid line (red) : $\Delta (U)$ inside the ribbon (middle site).  } \label{ribbon}
\end {figure}

In topological insulating states, the only free carriers is just the gapless edge states.  Beyond its linear dispersion, the spin of the edge state are locked with its momentum.
We study the zigzag ribbon structure so as to make clear the phase transition process of the edge states.
The zigzag ribbon structure is shown in Fig. 1 (b). It has infinite length
in the longitudinal direction (x direction) and finite width in the
transverse direction (y direction). The unit cell of the ribbon structure
is labeled by integer indices m and n. Due to the translational invariant,
after Fourier transformation along the x direction, the tight binding
Hamiltonian can be written as
\begin{equation}
H_{ribbon}=\int\frac{dk_{x}}{2\pi}\phi^{+}(k_{x})H_{ribbon}(k_{x})\phi(k_{x})
\end{equation}
where $\phi^{+}(k_{x})=[\cdots,a_{\uparrow}^{+}(k_{x}n),b_{\uparrow}^{+}(k_{x}n),\cdots,a_{\downarrow}^{+}(k_{x}n), \\ b_{\downarrow}^{+}(k_{x}n) ,\cdots]$
is the basis ($4N$ vector) with $n=1,2,\cdots,N$. Here $n$ is the row index and N is the width of ribbon, i.e.
the number of unit cell along the transverse cross section.
$H_{ribbon}(k_{x})$ is a $4N\times4N$ matrix
\begin{equation}
H_{ribbon}(k_{x})=\left(\begin{array}{cc}
H_{\uparrow}(k_{x}) & 0\\
0 & H_{\downarrow}(k_{x})\end{array}\right)
\end{equation}
where
$H_{\sigma=\uparrow\downarrow}(k_{x})$ are $2N\times2N$ tridiagonal matrices. Detail of the expression is given in appendix.
The attractive U interaction is still treated with mean field approximation
\begin{equation}
\begin{split}
H_{U}=&-\sum_{nk_{x}}[\Delta_{An}a_{\uparrow}^{+}(k_{x}n)a_{\downarrow}^{+}(-k_{x}n) \\ &+\Delta_{Bn}b^{+}(k_{x}n)b^{+}(-k_{x}n)] + h.c
\end{split}
\end{equation}
where $\Delta_{An}=\frac{|U|}{N_{r}}\sum_{k_{x}}\left\langle a_{\downarrow}(-k_{x}n)a_{\uparrow}(k_{x}n)\right\rangle $.
$\Delta_{An}$ ($\Delta_{Bn}$)
is the pairing potential on sublattice A (B) of row n. $N_{r}$ is the number of the unit cell of each row.
Therefore, with the basis in Nambu representation $\Psi^{+}(k_{x})=[\cdots,a_{\uparrow}^{+}(k_{x}n),b_{\uparrow}^{+}(k_{x}n),\cdots,a_{\downarrow}(-k_{x}n),b_{\downarrow}(-k_{x}n),\cdots]$,
the mean-field BCS Hamiltonian can be expressed as $H_{MF}=\int\frac{dk_{x}}{2\pi}\Psi^{+}(k_{x})H_{sc}(k_{x})\Psi(k_{x})$ where
\[
H_{sc}=\left(\begin{array}{cc}
H_{\uparrow}(k_{x}) & -\Delta_R\\
-\Delta_R^{*} & -H_{\downarrow}^{T}(-k_{x})\end{array}\right)\]
Here $\Delta_R$ is a diagonal matrix, the diagonal elements of which is
$(\ldots,\Delta_{An},\Delta_{Bn},\cdots)$.
Given $\Delta_{An}$, $\Delta_{Bn}$ and the chemical potential (i.e. the average electron density $n_e= N_e / [2N_r \cdot N]$), $H_{sc}$ can be diagonalized numerically and
we can get the energy dispersion and eigenfunction of the Bogoliubov quasiparticles.  Thus, $\Delta_{An}$ and $\Delta_{Bn}$ can be determined self-consistently. One thing should be noticed that since the appearance of the ribbon edge, $\Delta_{An}$ and $\Delta_{Bn}$ are no longer equivalent in the calculation of ribbon structure.  The results for half filling case are shown in Fig. 3.

In Fig. 3 (a), we see that below $U_c$ the gap $\Delta$ in the bulk is zero, and $\Delta$ around the edge is nonzero. It means that the superconductivity only appears near the edges and the bulk is still insulating when $U<U_c$. This is just the ESS (edge superconducting state) we mentioned above, which is the main results of this paper. Normally, the mechanism of edge superconductivity  mainly concerns the proximity effect which is not intrinsic. Here our results indicate the possibility of intrinsic ESS. Meanwhile, we note that surface  superconductivity in topological flat band system has been studied recently\cite{Volovik}. It implies that the ESS is a general property of the system with topological protected edge states.  When $U>U_c$, superconductivity also appears in the bulk, but the gap $\Delta$ near the edges is still larger than that in the bulk.

In Fig. 3 (b), we show the U dependence of the gap $\Delta$ at the edge and in the bulk respectively. Clearly, bulk superconductivity has a critical $U_c$ but edge superconductivity has not. Small U can immediately induce nonzero edge superconductivity. It is because that 1D Dirac fermion (linear dispersion of the edge state) has a constant density of state. As a comparison, we have demonstrated that the zero density of states of 2D Dirac fermion around the Dirac points results in the finite $U_c$ of the bulk superconductivity when gap is zero at half filling. According to the discussion above, an interesting inference is that 2D ESS (surface superconducting state) also exist in 3D TI but with a finite $U_{edge}$ at half filing.
Due to the appearance of the ESS in the attractive Kane-Mele-Hubbard model, the phase transition from topological insulating state to the superconducting state actually splits into two steps. Increasing the attractive U, the topological insulating state will immediately change into to ESS with any small U, and  when $U>U_c$, the the whole system becomes superconductor.
We also calculate the slightly doping case ($\mu \neq 0$ but still in the bulk gap) and the results are qualitatively similar.

\begin{figure}
\centering
\includegraphics[width=8.5cm]{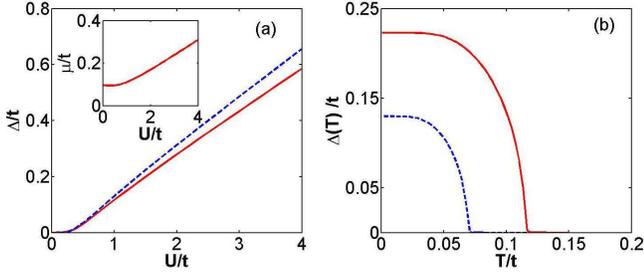}%
\caption{(Color online) (a) The self-consistent results of $\Delta (U)$ with the effective model of ESS (zero temperature). Dashed line (blue) :$\delta N =0$. Solid line (red) : $\delta N =0.15$.
We set $\hbar v_f=0.2 t\cdot a_1$ and momentum cutoff $k_c=\frac{\pi}{3a_1}$, where t is the hopping amplitude and $a_1=\sqrt{3}a_0$ is the 1D lattice constant of the tight binding model. The inner: mean field chemical potential $\mu$ versus $U$ for $\delta N =0.15$. (b) Temperature dependence of $\Delta$ for half filling case. Dashed line (blue): $U=1t$, Solid line (red): $U=1.5t$.
  } \label{zigzagribbon}
\end  {figure}

Finally, We study the effective model of the ESS. The effective model of the helical edge states is
$H_{0}= \hbar v_f \int dx  ( \Phi^+_{R\uparrow} i\partial_x \Phi_{R\uparrow} - \Phi^+_{L\downarrow} i\partial_x \Phi_{L\downarrow})$.
The attractive interaction is $H_U=-U \int dx \Phi^+_{R\uparrow} \Phi_{R\uparrow} \Phi^+_{L\downarrow} \Phi_{L\downarrow}$. Since the electron spin  is locked with its momentum, we  ignore the right-moving (left-moving) index R (L) in the following.
In momentum space, we have $H_0=\int \frac{dk}{2\pi} \psi^+H_0 (k) \psi$ where $\psi^+=[c^+_{k\uparrow}, c^+_{k\downarrow}]^T$, and
\[
H_{0}(k)=\left(\begin{array}{cc}
\hbar v_f k - \mu & 0\\
0 & -\hbar v_f k - \mu \end{array}\right)\]
The eigenvalues are $E_{\nu=\pm}=- \mu \pm \hbar v_f|k|$ with $\nu=\pm$ is the band index. Since not the spin but the helicity is  good quantum number here,  the upper ($\nu=+$) and lower ($\nu=-$) bands correspond to different helicity.
Based on the BCS mean field approximation, $H_U=-\int \frac{dk}{2\pi}  ( \Delta^*c_{-k\downarrow}c_{k\uparrow}+ \Delta c^+_{k\uparrow}c
^+_{-k \downarrow} )$ with $\Delta = U \int \frac{dk}{2\pi}<c_{-k\downarrow}c_{k\uparrow}>$. It should be noticed that when $k>0$ ($k<0$), $<c_{-k\downarrow}c_{k\uparrow}>$ indicate the pairing in upper band $\nu=+$ (lower band $\nu=-$).  Thus, $\Delta$ includes superconducting pairs in both bands.

It is convenient to express the superconducting Hamiltonian in the Nambu basis $\varphi^+=[c^+_{k\uparrow},c_{-k\downarrow}]$ but not in the band basis. We have
\[
H_{SC}=\left(\begin{array}{cc}
\hbar v_f k - \mu & -\Delta \\
-\Delta^* & -\hbar v_f k + \mu \end{array}\right)
\]
The energy dispersion of the Bogoliubov quasiparticle is $\epsilon_B (k) = \sqrt{\Delta^2 + (\hbar v_f k - \mu)^2}$. We  get the gap equation
\[\frac{1}{U}=\frac{1}{2}\int \frac{dk}{2\pi} \frac{\textrm{tanh}[\beta \epsilon_B(k)/2] }{\epsilon_B(k)}.\] In order to determine $\mu$ and $\Delta$, we need the number equation concerning the particle conservation
\[
\delta N = \int \frac{dk}{\pi} \{\nu^2+\frac{v_fk-\mu}{\epsilon_B(k)} f_F[\epsilon_B (k)]\}-N_{-}
\]
where $\nu^2=\frac{1}{2}[1-\frac{\hbar v_fk-\mu}{\epsilon_B(k)}]$, $f_F$ is the usual Fermi function and $N_{-}$ is the electron number of the filled band $\nu=-$. With above equations, we can determine the properties of the ESS with given $U$ and $\delta N$. The gap $\Delta (U)$ is calculated self-consistently for cases $\delta N=0$ (half filling) and $\delta N =0.15$ for example (See in Fig. 4). Since it is an effective model of lattice system, it is natural to use the hopping  t and 1D lattice constant $a_1=\sqrt{3}a_0$ as the unit. We can deduce the parameters, e.g. $v_f$, via fitting the tight binding band structure.
In Fig. 4 (a), the results show that when U is rather small the superconducting gap $\Delta$ is still nonzero which is qualitatively consistent  with the tight binding calculation. We also calculate the temperature dependence of the gap $\Delta$ in Fig. 4 (b).

The calculations above are only on the mean field level. However, the influence of fluctuation can not be ignored since it is very important and will kill the superconductivity in strictly 1D system. The ESS we studied here can actually be viewed as  a quasi-one-dimensional superconducting system which is similar as the ultrathin superconducting nanowire\cite{tinkham,sahu}. A key feature of quasi-one-dimensional superconducting nanowire is that thermally activated phase slip and quantum phase slip processes will induce finite resistance when $T<T_c$. Furthermore, because that the ESS  is not an isolated 1D system, it is possible that the coupling between the environments (e.g. the bulk or the substrate) may stabilize the superconductivity, which is like the case of carbon nanotube\cite{tang,takesue}.
The situations of ESS in real materials are complex and still an open question, but we propose that the conception of ESS
 could be experimentally realized and examined in atomic optical lattice system.  Recently, various schemes to realize topological
insulating state in optical lattice system have been proposed\cite{cooper,goldman,sarma,sinova}. And the attractive U
Hubbard model has been intensively studied in optical lattice system in order to investigate the BCS-BEC
crossover\cite{toschi,qchen,paredes}. So producing topological insulating state with attractive interaction is
straightforward in the optical lattice system . It means that it is possible to realize the conception of ESS in the
optical lattice system.

In summary, we investigate the phase transition from the topological insulating state to superconducting state in attractive Kane-Mele-Hubbard model with self-consistent mean field method. We clearly manifest the existence of the edge superconducting state which is superconducting at the edge but insulating in the bulk. The ESS results from the interplay between the attractive interaction and the special energy band of TI. Thus,  in contrast to the proximity effect induced edge superconducting state in TI,  the ESS we show here is intrinsic and is a general characteristic of the superconductivity of TI. In this model, increasing U, the ESS will occur immediately and when $U>U_c$ the whole system becomes superconductor. The critical $U_c$ of the bulk has been calculated. Due to the constant DOS of the edge state, there is no $U_c$ for the ESS. The effective model of the ESS has also been discussed. We  propose that the conception of ESS  could be experimentally realized in atomic optical system.

We acknowledge part of financial support from 
RGC GRF HKU 709211 and CRF HKUST3/09. YZ is supported by National Basic Research Program of China (973 Program,
No.2011CBA00103),
NSFC (No.11074218) and the Fundamental Research Funds for the Central
Universities in China.
JHG thanks Dr. Kai-Yu Yang, Prof. X. Dai and Prof. X. C. Xie for helpful discussion.



\renewcommand{\theequation}{A-\arabic{equation}}
\setcounter{equation}{0}  
\section*{APPENDIX}  

Here, we give the expression of $H_{\sigma=\uparrow \downarrow}$ in Eq. (5).

We have
\begin{equation*}
H_{\sigma}(k_{x})=\left(\begin{array}{ccccc}
H_{11}^{\sigma} & H_{12}^{\sigma} & 0 & \cdots & 0\\
H_{21}^{\uparrow\sigma} & H_{22}^{\sigma} & H_{23}^{\sigma} &  & \vdots\\
0 & \ddots & \ddots & \ddots & 0\\
\vdots &  & H_{N-1N-2}^{\sigma} & H_{N-1N-1}^{\sigma} & H_{N-1N}^{\sigma}\\
0 & \cdots & 0 & H_{NN-1}^{\sigma} & H_{NN}^{\sigma}\end{array}\right)
\end{equation*}
with
\begin{align}
&H_{nn}^{\sigma}=\left(\begin{array}{cc}
-\eta S_{k_{x}}-\mu & \chi_{k_{x}}\\                       \notag
\chi_{k_{x}}^{*} & \eta S_{k_{x}}-\mu\end{array}\right) \\ \notag
&H_{nn-1}^{\sigma}=\left(\begin{array}{cc}
\eta\bar{S}_{k_{x}} & -t\\
0 & -\bar{\eta S}_{k_{x}}\end{array}\right) \\
&H_{nn+1}^{\sigma}=\left(\begin{array}{cc}
\eta\bar{S}_{k_{x}}^{*} & 0\\
-t & -\eta\bar{S}_{k}^{*}\end{array}\right).                \notag
\end{align}
Here $ S_{k_{x}}=2\lambda_{so}sin(k_{x}a_1)$ and $\bar{S}_{k_{x}}=2\lambda_{so}sin(\frac{k_{x}a_1}{2})e^{i\frac{k_{x}a_1}{2}}$ concerning
the spin-orbit coupling; $\chi_{k}=-t\cdot(1+e^{-ik_{x}a_1})$ is related
with the next neighborhood hopping; $\eta=+1(-1)$ for spin up
(down). $a_1=\sqrt{3}a_0$ is the 1D lattice constant, i.e. the distance between  adjacent unit cells along the x direction [See in Fig. 1 (b)].



\begin{thebibliography}{}
\bibitem{cutoff}
Since it is just a model study which do not involve any concrete materials, we assume the Debye-like energy cutoff is just the band width in the tight binding calculation for simplicity. The calculation with small energy cutoff has also been done and we find the results are qualitatively similar.
\end{thebibliography}
\end{document}